\documentclass[aps,prb,twocolumn,floatfix]{revtex4}
\usepackage{times}
\usepackage{graphicx}
\usepackage{epsfig}
\usepackage{xcolor}
\usepackage{amsmath,bm}
\usepackage{amssymb}
\usepackage{chemformula} 
\usepackage[version=3]{mhchem}
\usepackage{nicefrac,xfrac}

\begin{document}
	
	\title{Disentangling competing interactions in disordered materials using interaction space modelling}
	
	\author{Ella M. Schmidt$^\ast$}
\affiliation
		{Faculty of Geosciences, MARUM and MAPEX, University of Bremen, Bremen, Germany}
		\author{Arkadiy Simonov}
\affiliation
	{Department of Materials, ETH Zurich, Zurich, Switzerland}
	
	\date{\today}

	\begin{abstract}
Understanding and manipulating the relationship between intentionally introduced disorder and material properties necessitates efficient characterization techniques. For example, single crystal diffuse scattering experiments provide insights into the driving forces behind local order phenomena.
In this work, we present a time- and resource-efficient approach based on mean field theory, that quantifies local interaction energies but unlike other techniques does not require computationally expensive supercell models.
The method is employed to quantify competing interactions in functionally disordered materials such as disordered rock salt cathode materials and Prussian blue analogs that share an underlying face-centred lattice.
\end{abstract}

\maketitle

	\section{Introduction}
	Crystalline materials exhibit long-range periodic order, yet they can also exhibit various forms of disorder. This disorder encompasses non-repeating variations in composition, bonding arrangements, molecular orientation, atomic displacements, and magnetic spins, which do not conform to a strictly ordered pattern \cite{simonov2020designing}. Disorder often displays a degree of short-range order (SRO) where certain configurations are locally favored but do not manifest as long-range ordered (LRO) arrangements. This phenomenon is exemplified in cases like geometric frustration in Ising triangular antiferromagnets \cite{Wannier_1950}, configurational degeneracy in cubic ice driven by hydrogen bonding \cite{Bernal_1933}, and long-period stacking phases in models like the anisotropic next-nearest neighbor interaction (ANNNI) model \cite{Bak_1982}.
	
	Disorder significantly influences the physical and chemical properties of materials, prompting interest in tuning disorder as a novel route to optimizing material properties \cite{simonov2020designing}. One of the principal challenges in contemporary structural science is understanding disordered structures, elucidating complex ordered structures, and delineating their structure-property relationships \cite{billinge2007problem, keen2015crystallography}. Complex disordered structures may arise from simple average structures and interactions \cite{Ziman_1979, Parsonage_1978}. Notable examples include \ch{Cu_{1-x}Au_{x}} alloys, disordered rock salt cathode materials \cite{ji2019hidden, szymanski_modeling_2023, chen2024exploring}, half-Heusler thermoelectric systems \cite{roth_simple_2020, roth_tuneable_2021}, relaxor ferroelectrics like \ch{BaTiO_3} \cite{senn2016emergence}, Prussian Blue analogs \cite{simonov2020hidden}, and metal-organic frameworks \cite{meekel2021correlated, meekel2023truchet}.
	
	In diffraction experiments, long-range periodic order manifests as sharp Bragg reflections, while deviations from LRO appear as broad and continuous diffuse scattering (DS). For materials with simple average structures, structure determination is routinely achieved through well established crystallographic methods, i.e. single crystal or powder diffraction. However, characterizing SRO is more complex and far from routine.
	
	The powder pair distribution function (PDF) is currently the most widely used method for experimentally discerning local ordering principles in functional and applied materials \cite{billinge2007problem, young2011applications}. The PDF, derived from the Fourier transform of powder diffraction patterns, presents a histogram of inter-atomic distances. While most features in the powder PDF are dominated by the average structure, correlated SRO induces subtle deviations in peak positions and intensities, complicating unambiguous interpretation due to overlapping interatomic distances \cite{klove2022machine}.
	
	Single crystal diffraction enables the collection of three-dimensional scattering data, offering distinct advantages over powder diffraction despite being experimentally more demanding. In single crystal diffraction, Bragg reflections and diffuse scattering can be more readily distinguished. Bragg reflections localize onto specific detector pixels, while diffuse scattering displays a broader and continuous intensity distribution, facilitating the separation of Bragg diffraction and diffuse scattering through punch and fill \cite{kobas2005structural} or outlier rejection algorithms \cite{weng_k-space_2020}. Conventional analysis methods interpret this separated DS atomistically or in terms of pairwise correlation parameters, often requiring a large set of parameters to describe disorder \cite{McGreevy_1988, Eremenko_2019, Goodwin_2019, weber2012three, Cowley_1950}.
	
	Currently, the most intuitive method for analyzing single-crystal DS is the three-dimensional difference pair distribution function (3D-$\Delta$PDF) \cite{weber2012three, roth2019solving}, which isolates DS from Bragg diffraction and maps local deviations from the average structure in a three-dimensional histogram of inter-atomic vectors. This approach enables direct interpretation of local correlations and facilitates quantitative refinement of disorder models in terms of pair-correlations \cite{simonov_experimental_2014, simonov2014yell}. However, its success hinges on the ability to reconstruct a continuous and gapless volume of three-dimensional reciprocal space during data processing, which can be challenging for complex experimental setups, e.g. in-situ and in-operado setups, or beam-sensitive materials.
	
	Access to the driving forces of the local order can be achieved through a direct Monte-Carlo (MC) simulation. In this simulation, the interaction potentials and/or their descriptive parameters are varied until the DS calculated from an atomistic model, generated using a MC simulation, reaches sufficient agreement with experimental data. This approach is implemented in the inverse Monte Carlo (IMC) method \cite{Weber_2005, Almarza_2003, Jain_2006b, DAlessandro_2011, Welberry:sh0147} and empirical potential structure refinement (EPSR) \cite{Soper_1996, Soper_2012}. While individual DS calculations are computationally efficient on modern computers, meaningful and complete refinement of diffuse scattering remains computationally demanding.
	
	In addition to these methods, a mean-field (MF) approach can be employed to analyze local order directly in interaction space \cite{Naya_1974, Nagai_1982, Derollez_1990, Descamps_1982, Paddison_2013, paddison2020scattering, schmidt2022efficient}. This approach circumvents the need for extensive atomistic modeling and provides direct insight into the underlying physics governing correlated disorder. In this manuscript, we demonstrate ultra-fast DS calculations and the application to compositional disordered rock salt structures (DRX). We begin by interpreting DRX DS, where local charge balance serves as the predominant driving force for the experimentally observed DS. Subsequently, we explore more complex structures where the competition between local charge balance and local centro-symmetry introduces greater complexity in the observed DS patterns.
	
\section{Results and discussion}
	\subsection{Local Charge Balance}
\begin{figure}
		\includegraphics[width=0.50\textwidth]{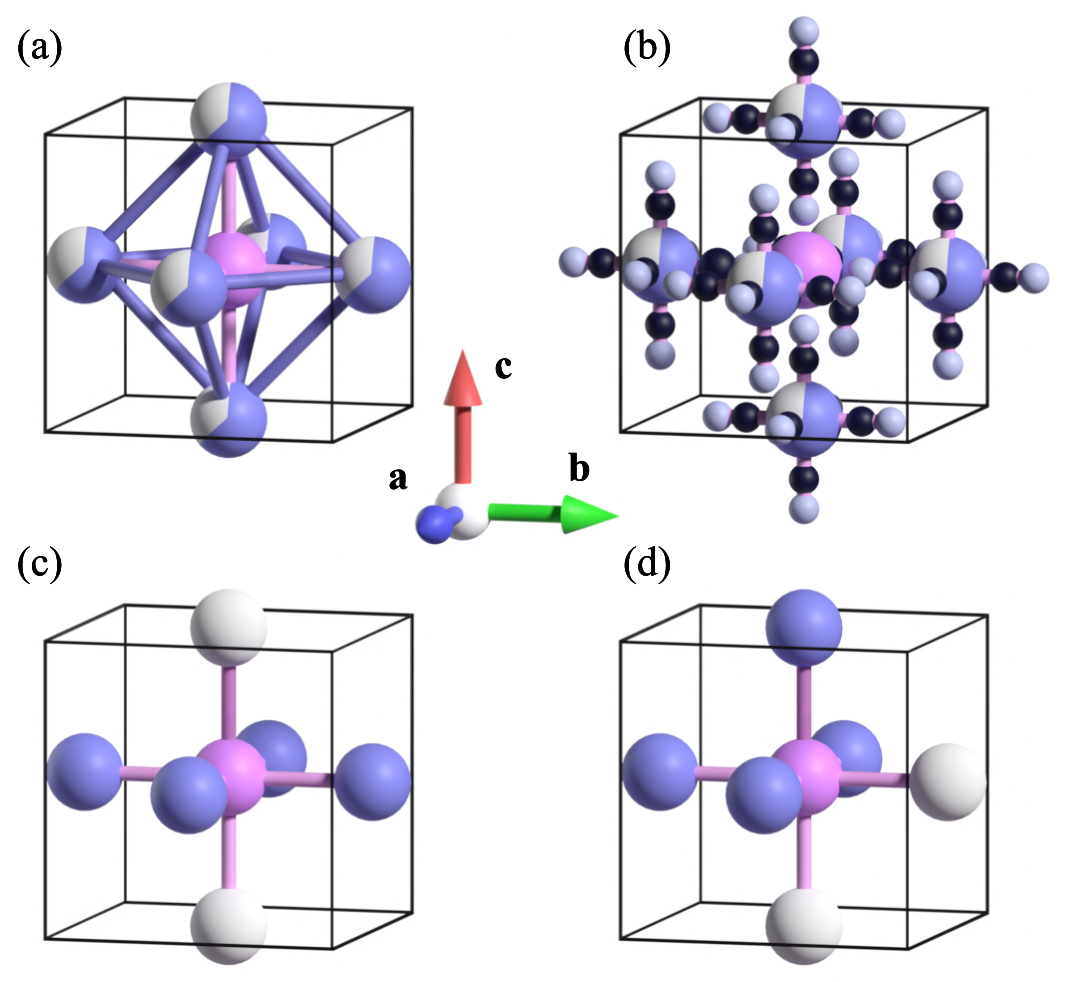}
		\caption{(a) Illustration of a DRX structure of a metal carbide \ch{MC_{1-x}}. Fully occupied and ordered \ch{M} in light pink, partially occupied \ch{C} in light purple. The box represents one unit-cell, bonds indicate the pair-wise inter-atomic vectors that need to be considered for local charge balance. Twelve nearest-neighbour vectors of type $\langle \frac{1}{2} \frac{1}{2}  0 \rangle$ are indicated in light purple and six next nearest neighbour vectors of type $\langle 1 0 0 \rangle$ are indicated in light pink. (b) Schematic illustration of the PBA average structure \ch{M[M'(CN)_6]}. Fully occupied \ch{M} atom in light pink, partially occupied \ch{M'(CN)_6}-octahedra in light blue. \ch{CN^-} groups in black and white. Black box indicates one unit cell. (c) Possible centro-symmetric vacancy arrangement, with occupied \ch{M'} atoms in lgiht purple and vacancies in white. (d) Possible acentric vacancy arrangement.}
		\label{fgr:Structures}
	\end{figure}

\begin{figure}
		\includegraphics[width=0.50\textwidth]{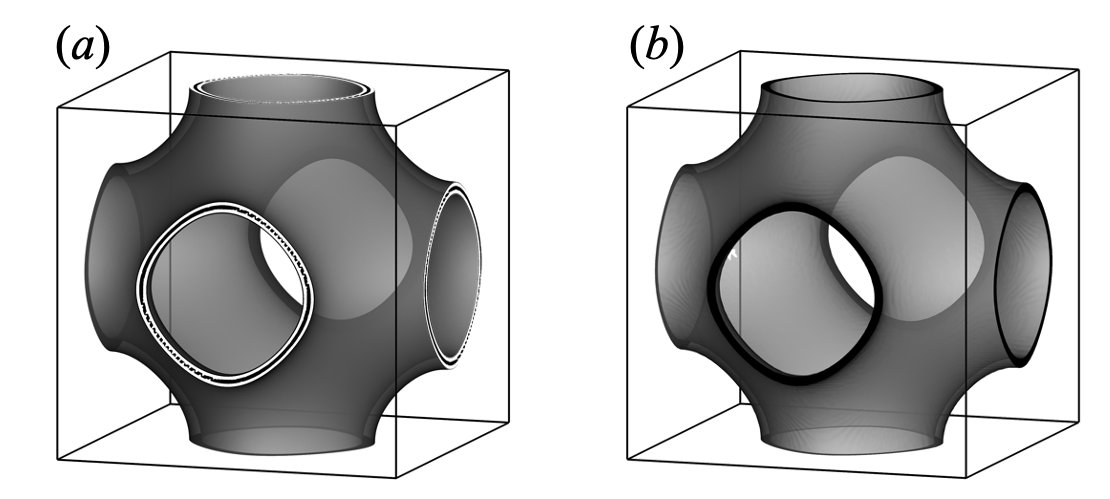}
		\caption{(a) Iso-surface rendering of Equation~\ref{eq:surf}. Box indicates the reciprocal space volume given by $0\le h,k,l \le 2$. (b) Iso-surface rendering using the MF approach for DRX structures with local charge balance. Box indicates the reciprocal space volume given by $0\le h,k,l \le 2$.  }
		\label{fgr:ChargeBalance}
	\end{figure}

	A diverse range of disordered solid-solution phases exhibit an average structure characteristic of compositional disordered rock salt (DRX) structures (see Figure~\ref{fgr:Structures}(a)), often displaying remarkably similar DS patterns (see Figure~\ref{fgr:ChargeBalance}(a) for an illustration) \cite{withers_disorder_2005}. Examples encompass non-stoichiometric transition metal carbides and nitrides \ch{MC_{1-x}} and \ch{MN_{1-x}} \cite{gusev_short-range_2006, sauvage_vacancy_1972, billingham_vacancy_1972}, as well as sub-stoichiometric early transition metal chalcogenides and their doped variants \cite{withers1994tem}.
	
	The DS can be characterized by a surface described by the equation\cite{withers_disorder_2005}:
	\begin{equation}
		\cos(\pi h) + \cos(\pi k) + \cos(\pi l) = 0
		\label{eq:surf}
	\end{equation}
	(see Figure~\ref{fgr:ChargeBalance}(a) for an illustration). Traditionally, solving this DS problem has relied on a cluster-expansion approach, which considers the local charge state of occupationally disordered ions surrounding a central fully ordered ion as the underlying mechanism for local ordering \cite{brunel1972determination, sauvage_vacancy_1972, de1977transition, withers_disorder_2005}. While effective for the mentioned systems, we leverage this analysis here to illustrate the power of the MF approach in DS analysis.
	
	The MF approach employs a pair-interaction Hamiltonian \cite{schmidt2022efficient, Naya_1974}:
	\begin{equation}
		\mathcal{H} = \frac{1}{2} \sum_{j} \sum_{k} \sum_{l=1}^{s} \sum_{m=1}^{s} \mu_{j}^{l} J^{lm}_{jk} \mu_{k}^{m},
		\label{eq:Hamilton1}
	\end{equation}
	where $j$ and $k$ sum over all unit cells in the crystal, and $l$ and $m$ sum over all $s$ disordered components. The variables $\mu_{j}^{l}$ take the value $1$ if the ion at site $j$ is occupied by species $l$, and $0$ otherwise; $J^{lm}_{jk}$ represents components of the pair-interaction Hamiltonian.
	
	Expressing $J^{lm}_{jk}$ as an $s$-dimensional matrix $\underline{\underline{J}}$ enables the diffuse scattering intensity to be described in terms of the Fourier-transformed pair-interaction Hamiltonian $\underline{\underline{J}}(\bm{H})$ \cite{schmidt2022efficient, Naya_1974}:
	\begin{equation}
		I(\bm{H}) \propto\mathrm{Tr}\left\{\underline{\underline{M}}\underline{\underline{F}}\left[\underline{\underline{1}} + \beta \underline{\underline{M}}\underline{\underline{J}}(\bm{H}) \right]^{-1}\right\},
		\label{eq:Hamilton2}
	\end{equation}
	where $\underline{\underline{M}}$ is an $s\times s$ matrix that encodes the average occupations of the disordered components. The elements of $\underline{\underline{M}}$ are given by:
	\[M_{ij} =m_i\delta_{ij} - m_i m_j \]
	where $m_i$ is the average occupation of species $i$. Here, $\beta = \frac{1}{k_{\mathrm{B}}T}$ denotes the inverse thermodynamic temperature, and $\underline{\underline{F}}$ represents the $s$-dimensional matrix of atomic form factors of the disordered components.
	
	The local order in these DRX is driven by local charge balance, i.e. the occupation of ions within the octahedron around the central ordered ion should resemble the average chemical composition as closely as possible\cite{withers_disorder_2005}, see Figure~\ref{fgr:Structures}(a) for an illustration. Fore example in the \ch{MC_{1-x}} system, each octahedron with a central \ch{M} atom is build by $6(1-x)$\ch{C} and $6x$ vacancies. As $6(1-x)$ and $6x$ are not necessarily integer numbers, configurations that are as close as possible to this average occupation are preferred.
	For translating this multi body interaction into pair-interactions, all inter-atomic vectors of the corners of the octahedron need to be considered. These are the twelve nearest-neighbour (NN) vectors of type $\langle \frac{1}{2} \frac{1}{2}  0 \rangle$ and six next-nearest-neighbour (NNN) vectors of type $\langle 1 0 0 \rangle$, as indicated in blue and purple in Figure~\ref{fgr:Structures}(a).
	To achieve local charge balance, like-wise pairs along these inter-atomic vectors are discouraged compared to random order, while unlike pairs are encouraged. This relationship is encapsulated in the pair-interaction matrix $\underline{\underline{J}}(\bm{H})$:
	\begin{equation}
		\begin{split}
			\underline{\underline{J}}(\bm{H}) = j_1 \underline{\underline{M}} \cdot \left( \right. & 2 \cos(\pi(h+k)) +  2 \cos(\pi(h-k))  \\ 
			+ & 2 \cos(\pi(k+l))  + 2  \cos(\pi(k-l)) \\  
			+  & 2 \cos(\pi(h+l)) + 2 \cos(\pi(h-l)) \\ 
			+ & \cos(2\pi h)  + \cos(2\pi k)  + \cos(2\pi l) \left. \right),
		\end{split}
		\label{eq:Jq}
	\end{equation}
	where $j_1$ is a proportionality constant.
	
	Equation \ref{eq:Hamilton2} can be effectively solved by selecting the proportionality constant $j_1$ in Equation \ref{eq:Jq} (here, $\beta \cdot j_1 = 0.75$) to satisfy the stability criterion:
	\begin{equation}
		\det \left[ \underline{\underline{1}} + \beta \underline{\underline{M}}\underline{\underline{J}}(\bm{H}) \right] \ge 0
		\label{eq:MFstability}
	\end{equation}
	for all $\bm{H}$ values.
	The resulting iso-surface of the diffuse scattering shown in Figure~\ref{fgr:ChargeBalance}(b) closely resembles the analytical solution derived from the cluster-expansion mechanism shown in Figure~\ref{fgr:ChargeBalance}(a) for comparison. This demonstrates the efficacy of the MF approach.

\subsection{Prussian blue analogs}
	\subsection{Prussian blue analogs}
	Systems where local charge balance is not the sole driving force for local order offer intriguing opportunities for tuning material properties. In particular, systems exhibiting intertwined competing interactions, such as local charge balance and local centro-symmetry, present promising avenues for such tuning. Notable examples include defective half-Heusler systems \cite{roth_simple_2020} and Prussian blue analogs (PBAs) \cite{simonov2020hidden}, which will be discussed here in the context of the MF approach.
	
	PBAs have a parent structure based on a cubic lattice, characterized by the idealized composition \ch{M[M'(CN)_6]}. In this structure, atoms of type \ch{M} and \ch{M'} (typically transition-metal cations) occupy a rock salt arrangement and are octahedrally coordinated by bridging cyanide ions (\ch{CN$^-$}) (see Figure~\ref{fgr:Structures}(b)). Here, we focus on the scenario described by Simonov et al. \cite{simonov2020hidden}, involving PBAs with a nominal composition of \ch{M^{II}[M'^{III}(CN)_6]_{2/3}}$\square_{1/3} \cdot x$\ch{H_2O}, where \ch{M} is a $2+$ and \ch{M'} is a $3+$ transition metal ion. To maintain charge balance, $1/3$ of \ch{M'} sites are unoccupied.

	By varying chemical species, crystal growth conditions, and post-synthesis treatments \cite{simonov2020hidden, kholina_metastable_2022}, PBAs can exhibit different local order states. The primary driving forces, as described by \citet{simonov2020hidden}, include (1) local charge balance and (2) a drive for local centro-symmetry, yielding the two distinct cis- and trans vacancy configurations shown in Figure~\ref{fgr:Structures}(c) and (d) respectively. The pair-interaction Hamiltonian by \citet{simonov2020hidden} incorporates both these forces and can be directly translated into the MF formalism:
	\begin{equation}
		\begin{split}
			\underline{\underline{J}}(\bm{H}) = j_1 \underline{\underline{M}} \cdot \left( \right. & 2 \cos(\pi(h+k)) +  2 \cos(\pi(h-k))  \\ 
			+ & 2 \cos(\pi(k+l))  + 2  \cos(\pi(k-l)) \\  
			+  & 2 \cos(\pi(h+l)) + 2 \cos(\pi(h-l)) \\ 
			+ & \cos(2\pi h)  + \cos(2\pi k)  + \cos(2\pi l) \left. \right) \\
			- j_2 \underline{\underline{M}} \cdot \left( \right. &\cos(2\pi h)  + \cos(2\pi k)  + \cos(2\pi l) \left. \right),
		\end{split}
		\label{eq:Jq_pba}
	\end{equation}
	where $j_1$ quantifies the drive for local charge balance and $j_2$ represents the drive for local centro-symmetry.
	
	The way that Equation~\ref{eq:Jq_pba} is set up, negative $j_2$ lead to locally acentric configurations, while positive $j_2$ lead to locally centro-symmetric configurations.
	The thermodynamic parameter $\beta$ and the absolute values of $j_1$ and $j_2$ are coupled, effectively transforming the problem into a two-parameter problem with free parameters $J' = j_1 / j_2$ and $T' = 1 / {\beta j_2}$ (compare \citet{simonov2020hidden}).
	
		\begin{figure}
		\includegraphics[width=0.5\textwidth]{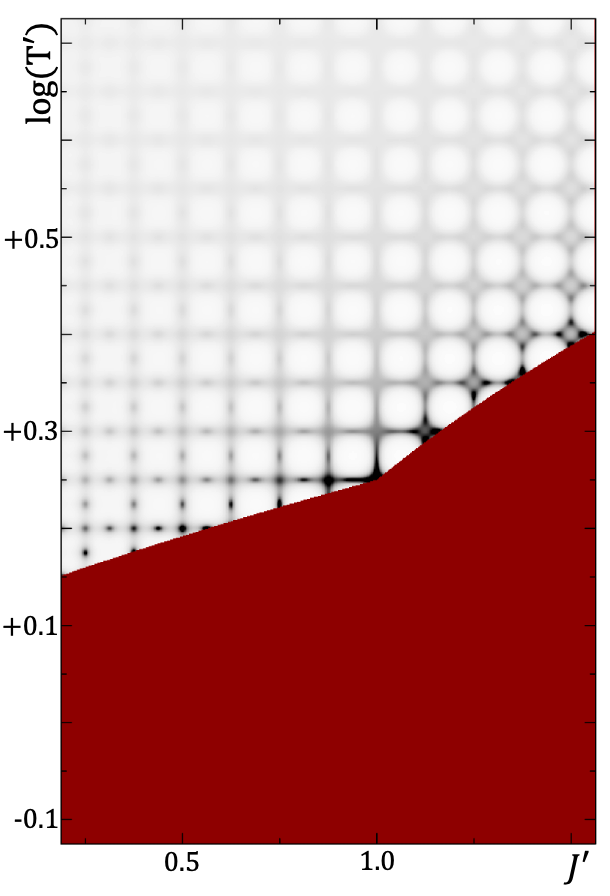}
		\caption{Diffuse scattering map of PBA as a function of local order parameters $J'$ and $T'$ generated with the MF approach and the Hamiltonian described in Equation~\ref{eq:Jq_pba}. Section in red indicates where the stability criterion in Equation~\ref{eq:MFstability} is not fulfilled.}
		\label{fgr:PBAMap}
	\end{figure}
	
	Figure~\ref{fgr:PBAMap} illustrates the DS map of PBAs generated using the MF approach, very closely resembling the map derived by \citet{simonov2020hidden} based on discrete tiles from a series of MC simulations. The MF approach can only access structures that lie within its stability regime (see Equation~\ref{eq:MFstability}), which admittedly here significantly limits the available coverage of the phase space compared to MC simulations, that have significantly less problems to access more ordered regimes.
	
	It is worth noting that, while direct MC simulations for generating disordered superstructures on modern computers are relatively efficient and recent advances in calculating single crystal DS (see e.g. \citet{paddison2020scattering}) have significantly decreased the computational cost of DS calculations, refining interaction potentials via MC simulations remains computationally demanding. This underscores the advantage of the presented MF approach for large datasets with potentially complex interactions, where refinement times can be reduced by several orders of magnitude.
	
\begin{figure}
		\includegraphics[width=0.49\textwidth]{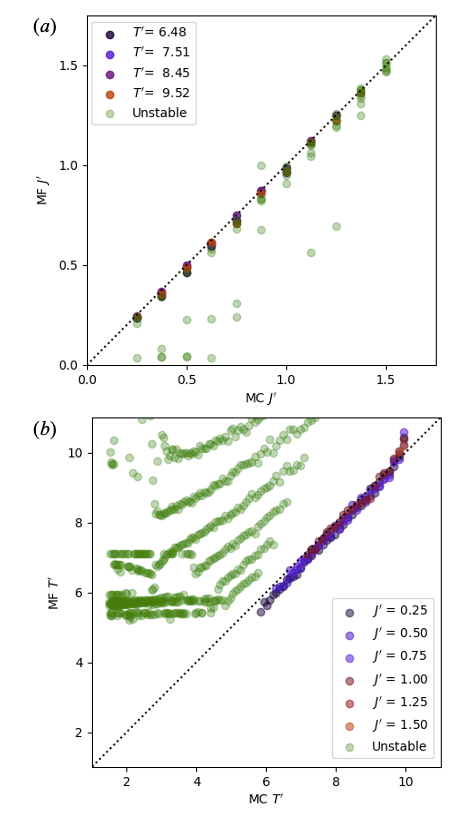}
		\caption{Refinement of pair-interaction using the MF algorithm on PBAs. (a) Comparison of MF refined $J'$ and $J'$ used in the MC simulation for five different MC temperatures $T'$. Configurations that do not fulfil the stability criterion in Equation~\ref{eq:MFstability} in the Monte Carlo Simulation are highlighted in green. Each dot corresponds to the refinement of one MC configuration. Black dotted line indicates perfect agreement.
			(b) Comparison of MF fit $1/j_2$ and the MC temperature for six different $J'$ used in the MC simulation. Configurations that do not fulfil the stability criterion in Equation~\ref{eq:MFstability} in the Monte Carlo Simulation are highlighted in green. Black dotted line indicates perfect agreement}
		\label{fgr:PBAFit}
	\end{figure}

	Furthermore, the MF approach is not limited to forward calculations of DS from a given pair-interaction Hamiltonian; it can also be applied in reverse. We demonstrate this by refining DS generated from coarse-grained MC simulations of PBAs using the Hamiltonian described by \citet{simonov2020hidden}. Refinement parameters $j_1$, $j_2$, and an overall scale are utilized in Equations \ref{eq:Hamilton2} and \ref{eq:Jq_pba}. The results of the MF refinement are compared to the original MC parameters in Figure~\ref{fgr:PBAFit}. 
	
	To accurately model the temperature dependence, one additional point must be taken into account, known as the reaction field in the magnetic analog \cite{paddison2023spinteract, PhysicsPhysiqueFizika.3.317, logan1995onsager}. Here, the temperature-dependent reaction field enforces self-consistency on the average spin length. In the structural MF approximation used here, this translates into a temperature-dependent self-consistency term that ensures the summed DS intensity within one Brillouin zone depends only on the chemical composition and is independent of the pair interactions. In practice, this is realized by re-normalizing the temperature parameter $1/j_2$ with a factor $\lambda = I_{j}/I_{j=0}$, where $I_{j}$ is the DS intensity summed in one Brillouin zone with the refined pair interactions $j_1$ and $j_2$, and $I_{j=0}$ is the DS intensity summed in one Brillouin zone with $j_1 = j_2 = 0$.
	
	Notably, within the stability field of the MF approximation (see Equation \ref{eq:MFstability}), the refined values for $J'$ and $T'$ very closely match the parameters used in the MC simulation as shown in Figure~\ref{fgr:PBAFit}. Slight deviations are expected, particularly for configurations outside the MF stability regime, where $J'$ remains mostly well approximated while the interaction strength is consistently underestimated.
	The observed behavior aligns with expectations, as the DS shape primarily reflects the relative strengths of competing interactions $j_1$ and $j_2$, whereas DS sharpness is influenced by absolute interaction strengths relative to the thermodynamic temperature $\beta$. As such, the MF approach offers an effective tool for DS analysis and parameter refinement in complex systems like PBAs.
 
\subsection{Disordered rock salt cathode materials}
The final system under consideration here is composed of disordered rock salt cathode materials, specifically \ch{Li_{1.2}Mn_{0.4}Ti_{0.4}O_2}  (LMTO) and \ch{Li_{1.2}Mn_{0.4}Zr_{0.4}O_2} (LMZO), as investigated by \citet{ji2019hidden}. As the previously discussed systems, these materials crystallize in a DRX structure. Due to the absence of large single crystals, electron diffraction was employed by \citet{ji2019hidden} to capture DS along selected zone axes. As a result, only limited reciprocal space data is available - a common challenge for many functional materials in their applied state\cite{schmidt2023quantitative}. 
	We want to use this example to demonstrate the robustness and reliability of the MF approach when limited reciprocal space data is available.
	
	\begin{figure}
		\includegraphics[width=0.49\textwidth]{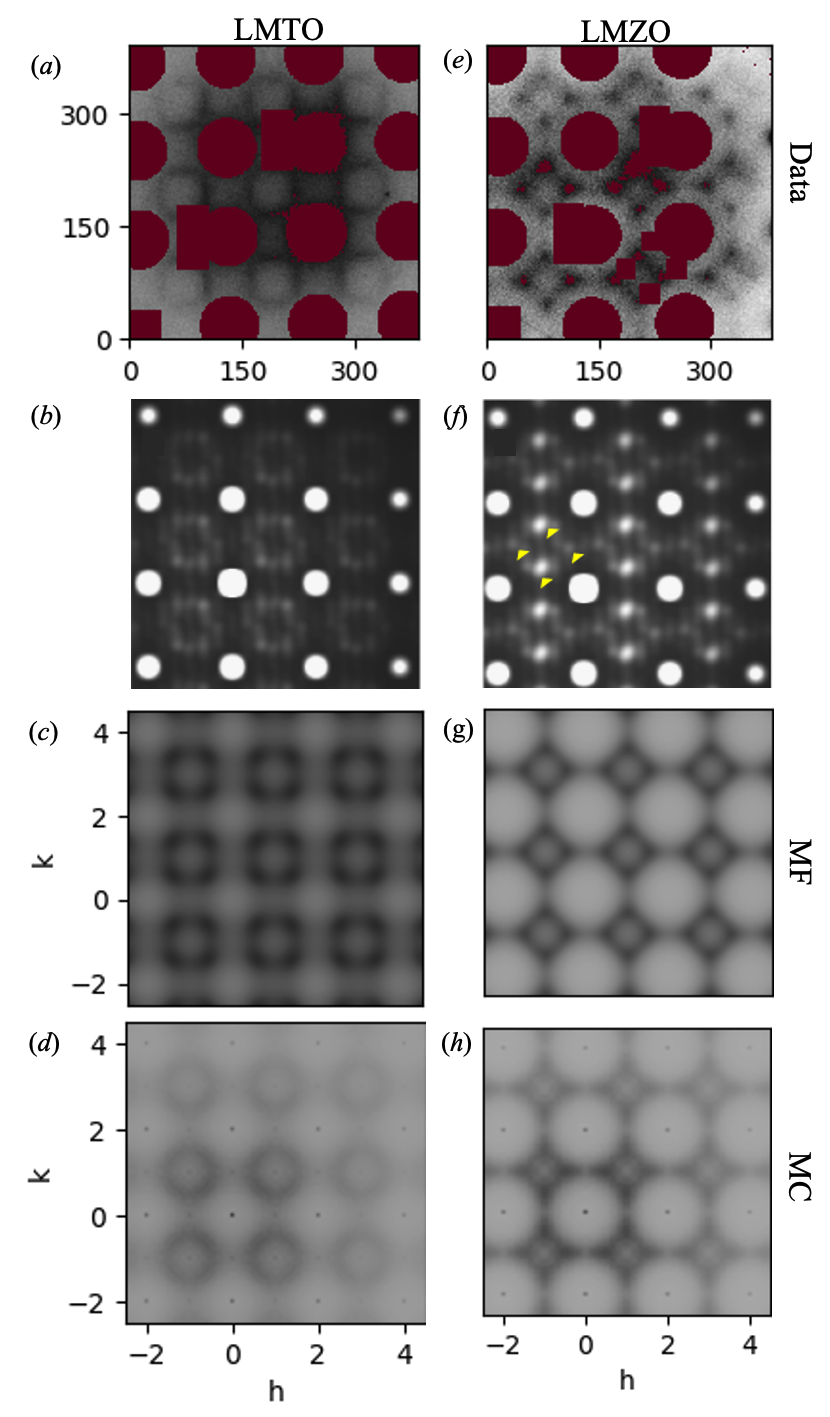}
		\caption{(a) Published LMTO diffuse scattering in the $[001]$ zone-axis by \citet{ji2019hidden}. Red areas indicate masking of Bragg reflections and over-exposed pixels, as well as labels in the published data. Axis units correspond to pixels. (b) Original fit by \citet{ji2019hidden}. (c) Results of the MF Fit using the Hamiltonian in Equation~\ref{eq:Jq_cathode}. (d) DS calculated from an MC simulation using the parameters determined by the MF fit.
			(e-h) Same as (a-d) for LMZO.  }
		\label{fgr:Cathode}
	\end{figure}

	LMTO and LMZO represent potential cathode materials for Li transport applications. In these materials, a well-connected Li network is vital, necessitating local Li clusters interconnected by nearest Li-Li neighbors. An optimal configuration therefore would not necessarily minimize the occurrence of likewise NN Li pairs.
	
	Similar to the previously discussed PBAs, the sharpness of diffuse scattering is primarily influenced by the absolute strength of interactions, while the shape is driven by the relative strengths of competing NN and NNN interactions, which we use here to describe the corresponding pair-interaction Hamiltonian:
	\begin{equation}
		\begin{split}
			\underline{\underline{J}}(\bm{H}) = j_1 \underline{\underline{M}} \cdot \left( \right. & 2 \cos(\pi(h+k)) +  2 \cos(\pi(h-k))  \\ 
			+ & 2 \cos(\pi(k+l))  + 2  \cos(\pi(k-l)) \\  
			+  & 2 \cos(\pi(h+l)) + 2 \cos(\pi(h-l))  \left. \right) \\
			+ j_2 \underline{\underline{M}} \cdot \left( \right. &\cos(2\pi h)  + \cos(2\pi k)  + \cos(2\pi l) \left. \right),
		\end{split}
		\label{eq:Jq_cathode}
	\end{equation}
	where $j_1$ quantifies the tendency to avoid like-nearest-neighbor Li pairs, and $j_2$ influences the relative strength of NN and NNN interactions. Note that the notation used for $j_1$ and $j_2$ in Equation~\ref{eq:Jq_pba} and Equation~\ref{eq:Jq_cathode} differ. For the PBA system the local charge balance term was chosen to align with the Hamiltonian used in \citet{simonov2020hidden}. For the DRX here, the Hamiltonian in Equation~\ref{eq:Jq_cathode} emphasizes the difference between NN and NNN interactions, similar to the analysis of \citet{ji2019hidden}.
	
	For the data analysis, we coarse-grain LMTO and LMZO as two-component systems, focusing on Li ($m_\mathrm{Li} = 0.6$) and the average of Mn/M' ($m_\mathrm{Mn/M'} = 0.4$) components within the structure.
	To disentangle the different Li-Mn, Li-M' and Mn-M' interactions effectively, additional complementary X-ray or neutron diffraction data would be necessary.
	
	To refine pair-interaction energies using the available electron diffraction data\cite{ji2019hidden} (three different zone axis: $[001]$, $[111]$ and $[110]$ for each of the two compounds are available), we assigned suitable $(hkl)$ values to each pixel in the published DS images, based on indicated Bragg reflections and gray-scale intensity conversions. 
	The refinement process involved optimizing eight parameters: three scale factors and three background parameters (one for each zone axis) along with $j_1$ and $j_2$ in Equation \ref{eq:Jq_cathode}. The resulting fit, illustrated in Figure~\ref{fgr:Cathode}(c) and (g), yielded refined parameters for $j_1$ and $j_2$, as shown in Table~\ref{tab:Cathode_j}.
	
	\begin{table}
		\begin{tabular}{|c|c|c|c|}
			\hline
			Material & $\beta \cdot j_1$  & $\beta \cdot j_2$ &  $j_1/j_2$ \\
			\hline
			LMTO & 0.2325  & 0.2436 &  1.2613 \\
			\hline
			LMZO&  0.8311 &  0.2933 & 0.2931  \\
			\hline
		\end{tabular}
		\caption{Refined $j_1$ and $j_2$ for the cathode materials LMTO and LMZO using the MF approach.}
		\label{tab:Cathode_j}
	\end{table}
	
	The experimentally observed DS is well reproduced by the MF fit.
	To test the reliability of the refined pair-interactions, we performed direct MC simulations with the refined parameters. 
	We utilized DISCUS\cite{neder2008diffuse} to simulate ten super cells of $20\times20\times20$ unit cells. The diffuse scattering in the $hk0$-layer was calculated using SCATTY\cite{paddison_ultrafast_2019} and is displayed in Figure~\ref{fgr:Cathode}(d) and (h). 
	The MC simulations reproduces the observed diffuse scattering even better than the MF fit which was used to derive the pair-interaction parameters. This confirms the reliability of our approach to the limited data coverage and demonstrates that the pair-interactions we derive here are better suited to reproduce the experimental diffuse scattering data than the cluster expansion Hamiltonian used in the Monte Carlo simulation by \citet{ji2019hidden} shown in Figure~\ref{fgr:Cathode}(b) and (f) respectively.
	
	\begin{figure}
		\includegraphics[width=8 cm]{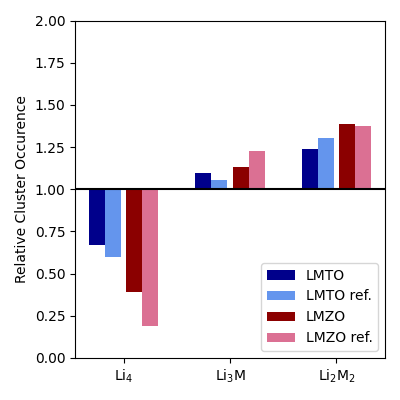}
		\caption{Relative cluster occurrence of different local clusters in LMTO and LMZO. Darker bars indicate results from our MC calculations using $j_1$ and $j_2$ determined with the MF approach, lighter bars compare to the results reported by \citet{ji2019hidden}. }
		\label{fgr:CathodeCluster}
	\end{figure}

	Moreover, the MC simulations provided insights into Li network statistics within the resulting super cells. The absolute value of $j_1$ significantly influenced the occurrence of local Li clusters, with larger $j_1$ values indicating stronger avoidance of like-NN Li pairs and thus lower probabilities of local Li clusters. The MF analysis suggest that in LMZO local cluster avoidance is stronger than in LMTO, in agreement with the findings of \citet{ji2019hidden}, who used cluster-expansion density functional theory (DFT) calculations to estimate ``relative cluster occurrence''. Figure~\ref{fgr:CathodeCluster} compares this ``relative cluster occurrence'' reported by \citet{ji2019hidden} to our MC simulations, leveraging simplified pair-interactions from the MF analysis.
	
	One last point that remains to be mentioned here is the interplay of competing NN and NNN pair-interactions. 
	While the absolute values of $j_1$ describes the occurrence of local Li clusters and the sharpness of the DS, the different ratios $j_1/j_2$ are what drives the different shapes of the observed DS scattering. Furthermore $j_1/j_2$ describes the relative strength of NN and NNN interactions, hence ultimately determining the connectivity between the different local Li clusters and hence the Li transport in the cathode materials.
 
\section{Conclusion and Outlook}
	This study highlights the efficacy of a simplified mean-field approximation for understanding complex disordered systems, demonstrated here on DRX structures. 
	Within the MF approach, the disorder is described in terms of very few but meaningful pair-interaction parameters, making the analysis viable for even the most complex systems and even when limited data is available. 
	The MF approach therefore provides insights into the key driving forces governing disorder.
	
	While the analysis demonstrated here is focused on DRX systems, the approach is versatile and applicable to complex materials systems. Generally, few competing interactions can drive complex disordered systems\cite{Ziman_1979,Goodwin_2019}. The use of MF analysis can disentangle these interactions and refine experimental data, avoiding large computational costs. For mangeitc systems, \citet{paddison2023spinteract} provides a user-friendly program that utilizes a similar MF approach to the one discussed here and provides detailed guidance on how to ensure meaningful refinement results.
	
	The robustness of the MF approach with limited data is particularly advantageous in advanced experimental setups - this characteristic underlines advantages as compared to other simple, direct analysis approaches such as 3D-$\Delta$PDFs.
	Examples where this may be an essential advantage include setups under pressure or electrical field, where limited reciprocal space coverage is expected as a result of the experimental peculiarities. 
	Another situation, where limited reciprocal space coverage may be an issue is when working with metastable or beam-sensitive materials, where the material alters faster than typical DS data acquisition times.
	
	Notably, we demonstrated that the stability criterion (Equation \ref{eq:MFstability}) serves as a guiding principle rather than a strict limitation of the approach presented here. Through refinement, the relative strengths of the underlying driving forces can be elucidated, although outside the stability regime the absolute strengths may be underestimated - clearly enlarging its applicability as compared to its first description by \citet{schmidt2022efficient}. 
	
	Looking ahead, several challenges remain to be addressed in the development and application of the MF approach to investigate complex disordered systems.
	One significant limitation of the current state of the MF approach arises when substitutional disorder is accompanied by pronounced displacement of scatterers from their average positions. While discrete displacements can be converted into different occupations, maintaining the discrete nature of degrees of freedom, it remains to be demonstrated in future research whether this approach is applicable and feasible for handling continuous displacements. This challenge is particularly pertinent for systems where continuous displacements play a crucial role in determining local disorder, e.g. cubic zirconia\cite{schmidt2023direct} or relaxor ferroelectrics\cite{pasciak2011diffuse}.
	Another important limitation requiring more detailed consideration is the treatment of systems driven by higher-order interactions that cannot be easily recast into pair interactions\cite{welberry1994interpretation}. Higher-order interactions may introduce additional complexity beyond pair-wise interactions, potentially necessitating alternative approaches for meaningful analysis and refinement.
	Moving forward, addressing these limitations will be essential for expanding the scope and applicability of the MF approach to a broader range of disordered materials and configurations. 
	However, despite these challenges, our study has demonstrated how the MF approach can be effectively utilized to investigate complex disordered systems, paving the way towards a more accessible refinement of meaningful interactions governing local disorder.

\section*{Acknowledgements}
	The authors thank Andrew L. Goodwin (Oxford University) and Joseph Paddison (Oak Ridge National Laboratory) for fruitful discussions and Paul Benjamin Klar (University of Bremen) for careful reading of the manuscript. We acknowledge the use of ChatGPT for manuscript editing.
\bibliography{MFRocksalt.bib}

\begin{thebibliography}{60}
\expandafter\ifx\csname natexlab\endcsname\relax\def\natexlab#1{#1}\fi
\expandafter\ifx\csname bibnamefont\endcsname\relax
  \def\bibnamefont#1{#1}\fi
\expandafter\ifx\csname bibfnamefont\endcsname\relax
  \def\bibfnamefont#1{#1}\fi
\expandafter\ifx\csname citenamefont\endcsname\relax
  \def\citenamefont#1{#1}\fi
\expandafter\ifx\csname url\endcsname\relax
  \def\url#1{\texttt{#1}}\fi
\expandafter\ifx\csname urlprefix\endcsname\relax\def\urlprefix{URL }\fi
\providecommand{\bibinfo}[2]{#2}
\providecommand{\eprint}[2][]{\url{#2}}

\bibitem[{\citenamefont{Simonov and Goodwin}(2020)}]{simonov2020designing}
\bibinfo{author}{\bibfnamefont{A.}~\bibnamefont{Simonov}} \bibnamefont{and}
  \bibinfo{author}{\bibfnamefont{A.~L.} \bibnamefont{Goodwin}},
  \bibinfo{journal}{Nature Reviews Chemistry} \textbf{\bibinfo{volume}{4}},
  \bibinfo{pages}{657} (\bibinfo{year}{2020}).

\bibitem[{\citenamefont{Wannier}(1950)}]{Wannier_1950}
\bibinfo{author}{\bibfnamefont{G.~H.} \bibnamefont{Wannier}},
  \bibinfo{journal}{Phys. Rev.} \textbf{\bibinfo{volume}{79}},
  \bibinfo{pages}{357} (\bibinfo{year}{1950}).

\bibitem[{\citenamefont{Bernal and Fowler}(1933)}]{Bernal_1933}
\bibinfo{author}{\bibfnamefont{J.~D.} \bibnamefont{Bernal}} \bibnamefont{and}
  \bibinfo{author}{\bibfnamefont{R.~H.} \bibnamefont{Fowler}},
  \bibinfo{journal}{J. Chem. Phys.} \textbf{\bibinfo{volume}{1}},
  \bibinfo{pages}{515} (\bibinfo{year}{1933}).

\bibitem[{\citenamefont{Bak}(1982)}]{Bak_1982}
\bibinfo{author}{\bibfnamefont{P.}~\bibnamefont{Bak}}, \bibinfo{journal}{Rep.
  Prog. Phys.} \textbf{\bibinfo{volume}{45}}, \bibinfo{pages}{587}
  (\bibinfo{year}{1982}).

\bibitem[{\citenamefont{Billinge and Levin}(2007)}]{billinge2007problem}
\bibinfo{author}{\bibfnamefont{S.~J.} \bibnamefont{Billinge}} \bibnamefont{and}
  \bibinfo{author}{\bibfnamefont{I.}~\bibnamefont{Levin}},
  \bibinfo{journal}{Science} \textbf{\bibinfo{volume}{316}},
  \bibinfo{pages}{561} (\bibinfo{year}{2007}).

\bibitem[{\citenamefont{Keen and Goodwin}(2015)}]{keen2015crystallography}
\bibinfo{author}{\bibfnamefont{D.~A.} \bibnamefont{Keen}} \bibnamefont{and}
  \bibinfo{author}{\bibfnamefont{A.~L.} \bibnamefont{Goodwin}},
  \bibinfo{journal}{Nature} \textbf{\bibinfo{volume}{521}},
  \bibinfo{pages}{303} (\bibinfo{year}{2015}).

\bibitem[{\citenamefont{Ziman}(1979)}]{Ziman_1979}
\bibinfo{author}{\bibfnamefont{J.~M.} \bibnamefont{Ziman}},
  \emph{\bibinfo{title}{Models of disorder. The theoretical physics of
  homogeneously disordered systems}} (\bibinfo{publisher}{Cambridge University
  Press}, \bibinfo{address}{Cambridge}, \bibinfo{year}{1979}).

\bibitem[{\citenamefont{Parsonage and Staveley}(1978)}]{Parsonage_1978}
\bibinfo{author}{\bibfnamefont{N.~G.} \bibnamefont{Parsonage}}
  \bibnamefont{and} \bibinfo{author}{\bibfnamefont{L.~A.~K.}
  \bibnamefont{Staveley}}, \emph{\bibinfo{title}{Disorder in Crystals}}
  (\bibinfo{publisher}{Clarendon Press}, \bibinfo{address}{Oxford},
  \bibinfo{year}{1978}).

\bibitem[{\citenamefont{Ji et~al.}(2019)\citenamefont{Ji, Urban, Kitchaev,
  Kwon, Artrith, Ophus, Huang, Cai, Shi, Kim et~al.}}]{ji2019hidden}
\bibinfo{author}{\bibfnamefont{H.}~\bibnamefont{Ji}},
  \bibinfo{author}{\bibfnamefont{A.}~\bibnamefont{Urban}},
  \bibinfo{author}{\bibfnamefont{D.~A.} \bibnamefont{Kitchaev}},
  \bibinfo{author}{\bibfnamefont{D.-H.} \bibnamefont{Kwon}},
  \bibinfo{author}{\bibfnamefont{N.}~\bibnamefont{Artrith}},
  \bibinfo{author}{\bibfnamefont{C.}~\bibnamefont{Ophus}},
  \bibinfo{author}{\bibfnamefont{W.}~\bibnamefont{Huang}},
  \bibinfo{author}{\bibfnamefont{Z.}~\bibnamefont{Cai}},
  \bibinfo{author}{\bibfnamefont{T.}~\bibnamefont{Shi}},
  \bibinfo{author}{\bibfnamefont{J.~C.} \bibnamefont{Kim}},
  \bibnamefont{et~al.}, \bibinfo{journal}{Nature communications}
  \textbf{\bibinfo{volume}{10}}, \bibinfo{pages}{1} (\bibinfo{year}{2019}).

\bibitem[{\citenamefont{Szymanski et~al.}(2023)\citenamefont{Szymanski, Lun,
  Liu, Self, Bartel, Nanda, Ouyang, and Ceder}}]{szymanski_modeling_2023}
\bibinfo{author}{\bibfnamefont{N.~J.} \bibnamefont{Szymanski}},
  \bibinfo{author}{\bibfnamefont{Z.}~\bibnamefont{Lun}},
  \bibinfo{author}{\bibfnamefont{J.}~\bibnamefont{Liu}},
  \bibinfo{author}{\bibfnamefont{E.~C.} \bibnamefont{Self}},
  \bibinfo{author}{\bibfnamefont{C.~J.} \bibnamefont{Bartel}},
  \bibinfo{author}{\bibfnamefont{J.}~\bibnamefont{Nanda}},
  \bibinfo{author}{\bibfnamefont{B.}~\bibnamefont{Ouyang}}, \bibnamefont{and}
  \bibinfo{author}{\bibfnamefont{G.}~\bibnamefont{Ceder}},
  \bibinfo{journal}{Chemistry of Materials} \textbf{\bibinfo{volume}{35}},
  \bibinfo{pages}{4922} (\bibinfo{year}{2023}).

\bibitem[{\citenamefont{Chen et~al.}(2024)\citenamefont{Chen, Leung, and
  Huang}}]{chen2024exploring}
\bibinfo{author}{\bibfnamefont{R.}~\bibnamefont{Chen}},
  \bibinfo{author}{\bibfnamefont{C.~L.~A.} \bibnamefont{Leung}},
  \bibnamefont{and} \bibinfo{author}{\bibfnamefont{C.}~\bibnamefont{Huang}},
  \bibinfo{journal}{Advanced Functional Materials} p. \bibinfo{pages}{2308165}
  (\bibinfo{year}{2024}).

\bibitem[{\citenamefont{Roth et~al.}(2020)\citenamefont{Roth, Zhu, and
  Iversen}}]{roth_simple_2020}
\bibinfo{author}{\bibfnamefont{N.}~\bibnamefont{Roth}},
  \bibinfo{author}{\bibfnamefont{T.}~\bibnamefont{Zhu}}, \bibnamefont{and}
  \bibinfo{author}{\bibfnamefont{B.~B.} \bibnamefont{Iversen}},
  \bibinfo{journal}{IUCrJ} \textbf{\bibinfo{volume}{7}}, \bibinfo{pages}{673}
  (\bibinfo{year}{2020}).

\bibitem[{\citenamefont{Roth et~al.}(2021)\citenamefont{Roth, Beyer, Fischer,
  Xia, Zhu, and Iversen}}]{roth_tuneable_2021}
\bibinfo{author}{\bibfnamefont{N.}~\bibnamefont{Roth}},
  \bibinfo{author}{\bibfnamefont{J.}~\bibnamefont{Beyer}},
  \bibinfo{author}{\bibfnamefont{K.~F.} \bibnamefont{Fischer}},
  \bibinfo{author}{\bibfnamefont{K.}~\bibnamefont{Xia}},
  \bibinfo{author}{\bibfnamefont{T.}~\bibnamefont{Zhu}}, \bibnamefont{and}
  \bibinfo{author}{\bibfnamefont{B.~B.} \bibnamefont{Iversen}},
  \bibinfo{journal}{IUCrJ} \textbf{\bibinfo{volume}{8}}, \bibinfo{pages}{695}
  (\bibinfo{year}{2021}).

\bibitem[{\citenamefont{Senn et~al.}(2016)\citenamefont{Senn, Keen, Lucas,
  Hriljac, and Goodwin}}]{senn2016emergence}
\bibinfo{author}{\bibfnamefont{M.}~\bibnamefont{Senn}},
  \bibinfo{author}{\bibfnamefont{D.}~\bibnamefont{Keen}},
  \bibinfo{author}{\bibfnamefont{T.}~\bibnamefont{Lucas}},
  \bibinfo{author}{\bibfnamefont{J.}~\bibnamefont{Hriljac}}, \bibnamefont{and}
  \bibinfo{author}{\bibfnamefont{A.}~\bibnamefont{Goodwin}},
  \bibinfo{journal}{Physical review letters} \textbf{\bibinfo{volume}{116}},
  \bibinfo{pages}{207602} (\bibinfo{year}{2016}).

\bibitem[{\citenamefont{Simonov et~al.}(2020)\citenamefont{Simonov,
  De~Baerdemaeker, Bostr{\"o}m, Rios~Gomez, Gray, Chernyshov, Bosak, B{\"u}rgi,
  and Goodwin}}]{simonov2020hidden}
\bibinfo{author}{\bibfnamefont{A.}~\bibnamefont{Simonov}},
  \bibinfo{author}{\bibfnamefont{T.}~\bibnamefont{De~Baerdemaeker}},
  \bibinfo{author}{\bibfnamefont{H.~L.} \bibnamefont{Bostr{\"o}m}},
  \bibinfo{author}{\bibfnamefont{M.~L.} \bibnamefont{Rios~Gomez}},
  \bibinfo{author}{\bibfnamefont{H.~J.} \bibnamefont{Gray}},
  \bibinfo{author}{\bibfnamefont{D.}~\bibnamefont{Chernyshov}},
  \bibinfo{author}{\bibfnamefont{A.}~\bibnamefont{Bosak}},
  \bibinfo{author}{\bibfnamefont{H.-B.} \bibnamefont{B{\"u}rgi}},
  \bibnamefont{and} \bibinfo{author}{\bibfnamefont{A.~L.}
  \bibnamefont{Goodwin}}, \bibinfo{journal}{Nature}
  \textbf{\bibinfo{volume}{578}}, \bibinfo{pages}{256} (\bibinfo{year}{2020}).

\bibitem[{\citenamefont{Meekel and Goodwin}(2021)}]{meekel2021correlated}
\bibinfo{author}{\bibfnamefont{E.~G.} \bibnamefont{Meekel}} \bibnamefont{and}
  \bibinfo{author}{\bibfnamefont{A.~L.} \bibnamefont{Goodwin}},
  \bibinfo{journal}{CrystEngComm} \textbf{\bibinfo{volume}{23}},
  \bibinfo{pages}{2915} (\bibinfo{year}{2021}).

\bibitem[{\citenamefont{Meekel et~al.}(2023)\citenamefont{Meekel, Schmidt,
  Cameron, Dharma, Windsor, Duyker, Minelli, Pope, Lepore, Slater
  et~al.}}]{meekel2023truchet}
\bibinfo{author}{\bibfnamefont{E.~G.} \bibnamefont{Meekel}},
  \bibinfo{author}{\bibfnamefont{E.~M.} \bibnamefont{Schmidt}},
  \bibinfo{author}{\bibfnamefont{L.~J.} \bibnamefont{Cameron}},
  \bibinfo{author}{\bibfnamefont{A.~D.} \bibnamefont{Dharma}},
  \bibinfo{author}{\bibfnamefont{H.~J.} \bibnamefont{Windsor}},
  \bibinfo{author}{\bibfnamefont{S.~G.} \bibnamefont{Duyker}},
  \bibinfo{author}{\bibfnamefont{A.}~\bibnamefont{Minelli}},
  \bibinfo{author}{\bibfnamefont{T.}~\bibnamefont{Pope}},
  \bibinfo{author}{\bibfnamefont{G.~O.} \bibnamefont{Lepore}},
  \bibinfo{author}{\bibfnamefont{B.}~\bibnamefont{Slater}},
  \bibnamefont{et~al.}, \bibinfo{journal}{Science}
  \textbf{\bibinfo{volume}{379}}, \bibinfo{pages}{357} (\bibinfo{year}{2023}).

\bibitem[{\citenamefont{Young and Goodwin}(2011)}]{young2011applications}
\bibinfo{author}{\bibfnamefont{C.~A.} \bibnamefont{Young}} \bibnamefont{and}
  \bibinfo{author}{\bibfnamefont{A.~L.} \bibnamefont{Goodwin}},
  \bibinfo{journal}{Journal of Materials Chemistry}
  \textbf{\bibinfo{volume}{21}}, \bibinfo{pages}{6464} (\bibinfo{year}{2011}).

\bibitem[{\citenamefont{Kl{\o}ve et~al.}(2023)\citenamefont{Kl{\o}ve, Sommer,
  Iversen, Hammer, and Dononelli}}]{klove2022machine}
\bibinfo{author}{\bibfnamefont{M.}~\bibnamefont{Kl{\o}ve}},
  \bibinfo{author}{\bibfnamefont{S.}~\bibnamefont{Sommer}},
  \bibinfo{author}{\bibfnamefont{B.~B.} \bibnamefont{Iversen}},
  \bibinfo{author}{\bibfnamefont{B.}~\bibnamefont{Hammer}}, \bibnamefont{and}
  \bibinfo{author}{\bibfnamefont{W.}~\bibnamefont{Dononelli}},
  \bibinfo{journal}{Advanced Materials} \textbf{\bibinfo{volume}{35}},
  \bibinfo{pages}{2208220} (\bibinfo{year}{2023}).

\bibitem[{\citenamefont{Kobas et~al.}(2005)\citenamefont{Kobas, Weber, and
  Steurer}}]{kobas2005structural}
\bibinfo{author}{\bibfnamefont{M.}~\bibnamefont{Kobas}},
  \bibinfo{author}{\bibfnamefont{T.}~\bibnamefont{Weber}}, \bibnamefont{and}
  \bibinfo{author}{\bibfnamefont{W.}~\bibnamefont{Steurer}},
  \bibinfo{journal}{Physical Review B} \textbf{\bibinfo{volume}{71}},
  \bibinfo{pages}{224206} (\bibinfo{year}{2005}).

\bibitem[{\citenamefont{Weng et~al.}(2020)\citenamefont{Weng, Dill, Martin,
  Whitfield, Hoffmann, and Ye}}]{weng_k-space_2020}
\bibinfo{author}{\bibfnamefont{J.}~\bibnamefont{Weng}},
  \bibinfo{author}{\bibfnamefont{E.~D.} \bibnamefont{Dill}},
  \bibinfo{author}{\bibfnamefont{J.~D.} \bibnamefont{Martin}},
  \bibinfo{author}{\bibfnamefont{R.}~\bibnamefont{Whitfield}},
  \bibinfo{author}{\bibfnamefont{C.}~\bibnamefont{Hoffmann}}, \bibnamefont{and}
  \bibinfo{author}{\bibfnamefont{F.}~\bibnamefont{Ye}},
  \bibinfo{journal}{Journal of Applied Crystallography}
  \textbf{\bibinfo{volume}{53}}, \bibinfo{pages}{159} (\bibinfo{year}{2020}).

\bibitem[{\citenamefont{McGreevy and Pusztai}(1988)}]{McGreevy_1988}
\bibinfo{author}{\bibfnamefont{R.~L.} \bibnamefont{McGreevy}} \bibnamefont{and}
  \bibinfo{author}{\bibfnamefont{L.}~\bibnamefont{Pusztai}},
  \bibinfo{journal}{Mol. Simul.} \textbf{\bibinfo{volume}{1}},
  \bibinfo{pages}{359} (\bibinfo{year}{1988}).

\bibitem[{\citenamefont{Eremenko et~al.}(2019)\citenamefont{Eremenko, Krayzman,
  Bosak, Playford, Chapman, Woicik, Ravel, and Levin}}]{Eremenko_2019}
\bibinfo{author}{\bibfnamefont{M.}~\bibnamefont{Eremenko}},
  \bibinfo{author}{\bibfnamefont{V.}~\bibnamefont{Krayzman}},
  \bibinfo{author}{\bibfnamefont{A.}~\bibnamefont{Bosak}},
  \bibinfo{author}{\bibfnamefont{H.~Y.} \bibnamefont{Playford}},
  \bibinfo{author}{\bibfnamefont{K.~W.} \bibnamefont{Chapman}},
  \bibinfo{author}{\bibfnamefont{J.~C.} \bibnamefont{Woicik}},
  \bibinfo{author}{\bibfnamefont{B.}~\bibnamefont{Ravel}}, \bibnamefont{and}
  \bibinfo{author}{\bibfnamefont{I.}~\bibnamefont{Levin}},
  \bibinfo{journal}{Nat. Commun.} \textbf{\bibinfo{volume}{10}},
  \bibinfo{pages}{2728} (\bibinfo{year}{2019}).

\bibitem[{\citenamefont{Goodwin}(2019)}]{Goodwin_2019}
\bibinfo{author}{\bibfnamefont{A.~L.} \bibnamefont{Goodwin}},
  \bibinfo{journal}{Nat. Commun.} \textbf{\bibinfo{volume}{10}},
  \bibinfo{pages}{4461} (\bibinfo{year}{2019}).

\bibitem[{\citenamefont{Weber and Simonov}(2012)}]{weber2012three}
\bibinfo{author}{\bibfnamefont{T.}~\bibnamefont{Weber}} \bibnamefont{and}
  \bibinfo{author}{\bibfnamefont{A.}~\bibnamefont{Simonov}},
  \bibinfo{journal}{Zeitschrift für Kristallographie - Crystalline Materials}
  \textbf{\bibinfo{volume}{227}}, \bibinfo{pages}{238} (\bibinfo{year}{2012}).

\bibitem[{\citenamefont{Cowley}(1950)}]{Cowley_1950}
\bibinfo{author}{\bibfnamefont{J.~M.} \bibnamefont{Cowley}},
  \bibinfo{journal}{Phys. Rev.} \textbf{\bibinfo{volume}{77}},
  \bibinfo{pages}{669} (\bibinfo{year}{1950}).

\bibitem[{\citenamefont{Roth and Iversen}(2019)}]{roth2019solving}
\bibinfo{author}{\bibfnamefont{N.}~\bibnamefont{Roth}} \bibnamefont{and}
  \bibinfo{author}{\bibfnamefont{B.~B.} \bibnamefont{Iversen}},
  \bibinfo{journal}{Acta Crystallographica Section A: Foundations and Advances}
  \textbf{\bibinfo{volume}{75}}, \bibinfo{pages}{465} (\bibinfo{year}{2019}).

\bibitem[{\citenamefont{Simonov
  et~al.}(2014{\natexlab{a}})\citenamefont{Simonov, Weber, and
  Steurer}}]{simonov_experimental_2014}
\bibinfo{author}{\bibfnamefont{A.}~\bibnamefont{Simonov}},
  \bibinfo{author}{\bibfnamefont{T.}~\bibnamefont{Weber}}, \bibnamefont{and}
  \bibinfo{author}{\bibfnamefont{W.}~\bibnamefont{Steurer}},
  \bibinfo{journal}{Journal of Applied Crystallography}
  \textbf{\bibinfo{volume}{47}}, \bibinfo{pages}{2011}
  (\bibinfo{year}{2014}{\natexlab{a}}).

\bibitem[{\citenamefont{Simonov
  et~al.}(2014{\natexlab{b}})\citenamefont{Simonov, Weber, and
  Steurer}}]{simonov2014yell}
\bibinfo{author}{\bibfnamefont{A.}~\bibnamefont{Simonov}},
  \bibinfo{author}{\bibfnamefont{T.}~\bibnamefont{Weber}}, \bibnamefont{and}
  \bibinfo{author}{\bibfnamefont{W.}~\bibnamefont{Steurer}},
  \bibinfo{journal}{Journal of Applied Crystallography}
  \textbf{\bibinfo{volume}{47}}, \bibinfo{pages}{1146}
  (\bibinfo{year}{2014}{\natexlab{b}}).

\bibitem[{\citenamefont{Weber}(2005)}]{Weber_2005}
\bibinfo{author}{\bibfnamefont{T.}~\bibnamefont{Weber}},
  \bibinfo{journal}{Zeitschrift f{\"u}r Kristallographie-Crystalline Materials}
  \textbf{\bibinfo{volume}{220}}, \bibinfo{pages}{1099} (\bibinfo{year}{2005}).

\bibitem[{\citenamefont{Almarza and Lomba}(2003)}]{Almarza_2003}
\bibinfo{author}{\bibfnamefont{N.~G.} \bibnamefont{Almarza}} \bibnamefont{and}
  \bibinfo{author}{\bibfnamefont{E.}~\bibnamefont{Lomba}},
  \bibinfo{journal}{Phys. Rev. E} \textbf{\bibinfo{volume}{68}},
  \bibinfo{pages}{011202} (\bibinfo{year}{2003}).

\bibitem[{\citenamefont{Jain et~al.}(2006)\citenamefont{Jain, Garde, and
  Kumar}}]{Jain_2006b}
\bibinfo{author}{\bibfnamefont{S.}~\bibnamefont{Jain}},
  \bibinfo{author}{\bibfnamefont{S.}~\bibnamefont{Garde}}, \bibnamefont{and}
  \bibinfo{author}{\bibfnamefont{S.~K.} \bibnamefont{Kumar}},
  \bibinfo{journal}{Ind. Eng. Chem. Res.} \textbf{\bibinfo{volume}{45}},
  \bibinfo{pages}{5614} (\bibinfo{year}{2006}).

\bibitem[{\citenamefont{D'Alessandro}(2011)}]{DAlessandro_2011}
\bibinfo{author}{\bibfnamefont{M.}~\bibnamefont{D'Alessandro}},
  \bibinfo{journal}{Phys. Rev. E} \textbf{\bibinfo{volume}{84}},
  \bibinfo{pages}{041130} (\bibinfo{year}{2011}).

\bibitem[{\citenamefont{Welberry}(2001)}]{Welberry:sh0147}
\bibinfo{author}{\bibfnamefont{T.~R.} \bibnamefont{Welberry}},
  \bibinfo{journal}{Acta Crystallographica Section A}
  \textbf{\bibinfo{volume}{57}}, \bibinfo{pages}{244} (\bibinfo{year}{2001}).

\bibitem[{\citenamefont{Soper}(1996)}]{Soper_1996}
\bibinfo{author}{\bibfnamefont{A.~K.} \bibnamefont{Soper}},
  \bibinfo{journal}{Chem. Phys.} \textbf{\bibinfo{volume}{202}},
  \bibinfo{pages}{295} (\bibinfo{year}{1996}).

\bibitem[{\citenamefont{Soper}(2012)}]{Soper_2012}
\bibinfo{author}{\bibfnamefont{A.~K.} \bibnamefont{Soper}},
  \bibinfo{journal}{Mol. Simul.} \textbf{\bibinfo{volume}{38}},
  \bibinfo{pages}{1171} (\bibinfo{year}{2012}).

\bibitem[{\citenamefont{Naya}(1974)}]{Naya_1974}
\bibinfo{author}{\bibfnamefont{S.}~\bibnamefont{Naya}}, \bibinfo{journal}{J.
  Phys. Soc. Jpn.} \textbf{\bibinfo{volume}{37}}, \bibinfo{pages}{340}
  (\bibinfo{year}{1974}).

\bibitem[{\citenamefont{Nagai}(1982)}]{Nagai_1982}
\bibinfo{author}{\bibfnamefont{K.}~\bibnamefont{Nagai}}, \bibinfo{journal}{J.
  Phys. Soc. Jpn.} \textbf{\bibinfo{volume}{51}}, \bibinfo{pages}{4015}
  (\bibinfo{year}{1982}).

\bibitem[{\citenamefont{Derollez et~al.}(1990)\citenamefont{Derollez, Lefebvre,
  and Descamps}}]{Derollez_1990}
\bibinfo{author}{\bibfnamefont{P.}~\bibnamefont{Derollez}},
  \bibinfo{author}{\bibfnamefont{J.}~\bibnamefont{Lefebvre}}, \bibnamefont{and}
  \bibinfo{author}{\bibfnamefont{M.}~\bibnamefont{Descamps}},
  \bibinfo{journal}{J. Phys.: Condens. Matter} \textbf{\bibinfo{volume}{2}},
  \bibinfo{pages}{9975} (\bibinfo{year}{1990}).

\bibitem[{\citenamefont{Descamps}(1982)}]{Descamps_1982}
\bibinfo{author}{\bibfnamefont{M.}~\bibnamefont{Descamps}},
  \bibinfo{journal}{J. Phys. C: Solid State Phys.}
  \textbf{\bibinfo{volume}{15}}, \bibinfo{pages}{7265} (\bibinfo{year}{1982}).

\bibitem[{\citenamefont{Paddison et~al.}(2013)\citenamefont{Paddison, Stewart,
  Manuel, Courtois, McIntyre, Rainford, and Goodwin}}]{Paddison_2013}
\bibinfo{author}{\bibfnamefont{J.~A.~M.} \bibnamefont{Paddison}},
  \bibinfo{author}{\bibfnamefont{J.~R.} \bibnamefont{Stewart}},
  \bibinfo{author}{\bibfnamefont{P.}~\bibnamefont{Manuel}},
  \bibinfo{author}{\bibfnamefont{P.}~\bibnamefont{Courtois}},
  \bibinfo{author}{\bibfnamefont{G.~J.} \bibnamefont{McIntyre}},
  \bibinfo{author}{\bibfnamefont{B.~D.} \bibnamefont{Rainford}},
  \bibnamefont{and} \bibinfo{author}{\bibfnamefont{A.~L.}
  \bibnamefont{Goodwin}}, \bibinfo{journal}{Phys. Rev. Lett.}
  \textbf{\bibinfo{volume}{110}}, \bibinfo{pages}{267207}
  (\bibinfo{year}{2013}).

\bibitem[{\citenamefont{Paddison}(2020)}]{paddison2020scattering}
\bibinfo{author}{\bibfnamefont{J.~A.} \bibnamefont{Paddison}},
  \bibinfo{journal}{Physical Review Letters} \textbf{\bibinfo{volume}{125}},
  \bibinfo{pages}{247202} (\bibinfo{year}{2020}).

\bibitem[{\citenamefont{Schmidt et~al.}(2022)\citenamefont{Schmidt, Bulled, and
  Goodwin}}]{schmidt2022efficient}
\bibinfo{author}{\bibfnamefont{E.~M.} \bibnamefont{Schmidt}},
  \bibinfo{author}{\bibfnamefont{J.~M.} \bibnamefont{Bulled}},
  \bibnamefont{and} \bibinfo{author}{\bibfnamefont{A.~L.}
  \bibnamefont{Goodwin}}, \bibinfo{journal}{IUCrJ}
  \textbf{\bibinfo{volume}{9}}, \bibinfo{pages}{21} (\bibinfo{year}{2022}).

\bibitem[{\citenamefont{Withers}(2005)}]{withers_disorder_2005}
\bibinfo{author}{\bibfnamefont{R.~L.} \bibnamefont{Withers}},
  \bibinfo{journal}{Zeitschrift fur Kristallographie}
  \textbf{\bibinfo{volume}{220}}, \bibinfo{pages}{1027} (\bibinfo{year}{2005}).

\bibitem[{\citenamefont{Gusev}(2006)}]{gusev_short-range_2006}
\bibinfo{author}{\bibfnamefont{A.~I.} \bibnamefont{Gusev}},
  \bibinfo{journal}{Physics-Uspekhi} \textbf{\bibinfo{volume}{49}},
  \bibinfo{pages}{693} (\bibinfo{year}{2006}).

\bibitem[{\citenamefont{Sauvage and Parthé}(1972)}]{sauvage_vacancy_1972}
\bibinfo{author}{\bibfnamefont{M.}~\bibnamefont{Sauvage}} \bibnamefont{and}
  \bibinfo{author}{\bibfnamefont{E.}~\bibnamefont{Parthé}},
  \bibinfo{journal}{Acta Crystallographica Section A}
  \textbf{\bibinfo{volume}{28}}, \bibinfo{pages}{607} (\bibinfo{year}{1972}).

\bibitem[{\citenamefont{Billingham et~al.}(1972)\citenamefont{Billingham, Bell,
  and Lewis}}]{billingham_vacancy_1972}
\bibinfo{author}{\bibfnamefont{J.}~\bibnamefont{Billingham}},
  \bibinfo{author}{\bibfnamefont{P.~S.} \bibnamefont{Bell}}, \bibnamefont{and}
  \bibinfo{author}{\bibfnamefont{M.~H.} \bibnamefont{Lewis}},
  \bibinfo{journal}{Acta Crystallographica Section A}
  \textbf{\bibinfo{volume}{28}}, \bibinfo{pages}{602} (\bibinfo{year}{1972}).

\bibitem[{\citenamefont{Withers et~al.}(1994)\citenamefont{Withers, Otero-Diaz,
  and Thompson}}]{withers1994tem}
\bibinfo{author}{\bibfnamefont{R.}~\bibnamefont{Withers}},
  \bibinfo{author}{\bibfnamefont{L.}~\bibnamefont{Otero-Diaz}},
  \bibnamefont{and} \bibinfo{author}{\bibfnamefont{J.}~\bibnamefont{Thompson}},
  \bibinfo{journal}{Journal of Solid State Chemistry}
  \textbf{\bibinfo{volume}{111}}, \bibinfo{pages}{283} (\bibinfo{year}{1994}).

\bibitem[{\citenamefont{Brunel et~al.}(1972)\citenamefont{Brunel, De~Bergevin,
  and Gondrand}}]{brunel1972determination}
\bibinfo{author}{\bibfnamefont{M.}~\bibnamefont{Brunel}},
  \bibinfo{author}{\bibfnamefont{F.}~\bibnamefont{De~Bergevin}},
  \bibnamefont{and} \bibinfo{author}{\bibfnamefont{M.}~\bibnamefont{Gondrand}},
  \bibinfo{journal}{Journal of Physics and Chemistry of Solids}
  \textbf{\bibinfo{volume}{33}}, \bibinfo{pages}{1927} (\bibinfo{year}{1972}).

\bibitem[{\citenamefont{De~Ridder et~al.}(1977)\citenamefont{De~Ridder,
  Van~Tendeloo, Van~Dyck, and Amelinckx}}]{de1977transition}
\bibinfo{author}{\bibfnamefont{R.}~\bibnamefont{De~Ridder}},
  \bibinfo{author}{\bibfnamefont{G.}~\bibnamefont{Van~Tendeloo}},
  \bibinfo{author}{\bibfnamefont{D.}~\bibnamefont{Van~Dyck}}, \bibnamefont{and}
  \bibinfo{author}{\bibfnamefont{S.}~\bibnamefont{Amelinckx}},
  \bibinfo{journal}{Le Journal de Physique Colloques}
  \textbf{\bibinfo{volume}{38}}, \bibinfo{pages}{C7} (\bibinfo{year}{1977}).

\bibitem[{\citenamefont{Kholina et~al.}(2022)\citenamefont{Kholina, Dössegger,
  Weber, and Simonov}}]{kholina_metastable_2022}
\bibinfo{author}{\bibfnamefont{Y.}~\bibnamefont{Kholina}},
  \bibinfo{author}{\bibfnamefont{J.}~\bibnamefont{Dössegger}},
  \bibinfo{author}{\bibfnamefont{M.~C.} \bibnamefont{Weber}}, \bibnamefont{and}
  \bibinfo{author}{\bibfnamefont{A.}~\bibnamefont{Simonov}},
  \bibinfo{journal}{Acta Crystallographica Section B Structural Science,
  Crystal Engineering and Materials} \textbf{\bibinfo{volume}{78}}
  (\bibinfo{year}{2022}).

\bibitem[{\citenamefont{Paddison}(2023)}]{paddison2023spinteract}
\bibinfo{author}{\bibfnamefont{J.~A.} \bibnamefont{Paddison}},
  \bibinfo{journal}{Journal of Physics: Condensed Matter}
  \textbf{\bibinfo{volume}{35}}, \bibinfo{pages}{495802}
  (\bibinfo{year}{2023}).

\bibitem[{\citenamefont{Brout and Thomas}(1967)}]{PhysicsPhysiqueFizika.3.317}
\bibinfo{author}{\bibfnamefont{R.}~\bibnamefont{Brout}} \bibnamefont{and}
  \bibinfo{author}{\bibfnamefont{H.}~\bibnamefont{Thomas}},
  \bibinfo{journal}{Physics Physique Fizika} \textbf{\bibinfo{volume}{3}},
  \bibinfo{pages}{317} (\bibinfo{year}{1967}),
  \urlprefix\url{https://link.aps.org/doi/10.1103/PhysicsPhysiqueFizika.3.317}.

\bibitem[{\citenamefont{Logan et~al.}(1995)\citenamefont{Logan, Szczech, and
  Tusch}}]{logan1995onsager}
\bibinfo{author}{\bibfnamefont{D.}~\bibnamefont{Logan}},
  \bibinfo{author}{\bibfnamefont{Y.}~\bibnamefont{Szczech}}, \bibnamefont{and}
  \bibinfo{author}{\bibfnamefont{M.}~\bibnamefont{Tusch}},
  \bibinfo{journal}{Europhysics Letters} \textbf{\bibinfo{volume}{30}},
  \bibinfo{pages}{307} (\bibinfo{year}{1995}).

\bibitem[{\citenamefont{Schmidt
  et~al.}(2023{\natexlab{a}})\citenamefont{Schmidt, Klar, Krysiak, Svora,
  Goodwin, and Palatinus}}]{schmidt2023quantitative}
\bibinfo{author}{\bibfnamefont{E.~M.} \bibnamefont{Schmidt}},
  \bibinfo{author}{\bibfnamefont{P.~B.} \bibnamefont{Klar}},
  \bibinfo{author}{\bibfnamefont{Y.}~\bibnamefont{Krysiak}},
  \bibinfo{author}{\bibfnamefont{P.}~\bibnamefont{Svora}},
  \bibinfo{author}{\bibfnamefont{A.~L.} \bibnamefont{Goodwin}},
  \bibnamefont{and}
  \bibinfo{author}{\bibfnamefont{L.}~\bibnamefont{Palatinus}},
  \bibinfo{journal}{Nature Communications} \textbf{\bibinfo{volume}{14}},
  \bibinfo{pages}{6512} (\bibinfo{year}{2023}{\natexlab{a}}).

\bibitem[{\citenamefont{Neder and Proffen}(2008)}]{neder2008diffuse}
\bibinfo{author}{\bibfnamefont{R.~B.} \bibnamefont{Neder}} \bibnamefont{and}
  \bibinfo{author}{\bibfnamefont{T.}~\bibnamefont{Proffen}},
  \emph{\bibinfo{title}{Diffuse Scattering and Defect Structure Simulations: A
  cook book using the program DISCUS}}, vol.~\bibinfo{volume}{11}
  (\bibinfo{publisher}{OUP Oxford}, \bibinfo{year}{2008}).

\bibitem[{\citenamefont{Paddison}(2019)}]{paddison_ultrafast_2019}
\bibinfo{author}{\bibfnamefont{J.~A.} \bibnamefont{Paddison}},
  \bibinfo{journal}{Acta Crystallographica Section A: Foundations and Advances}
  \textbf{\bibinfo{volume}{75}}, \bibinfo{pages}{14} (\bibinfo{year}{2019}),
  ISSN \bibinfo{issn}{20532733}.

\bibitem[{\citenamefont{Schmidt
  et~al.}(2023{\natexlab{b}})\citenamefont{Schmidt, Neder, Martin, Minelli,
  Lem{\'e}e, and Goodwin}}]{schmidt2023direct}
\bibinfo{author}{\bibfnamefont{E.~M.} \bibnamefont{Schmidt}},
  \bibinfo{author}{\bibfnamefont{R.~B.} \bibnamefont{Neder}},
  \bibinfo{author}{\bibfnamefont{J.~D.} \bibnamefont{Martin}},
  \bibinfo{author}{\bibfnamefont{A.}~\bibnamefont{Minelli}},
  \bibinfo{author}{\bibfnamefont{M.-H.} \bibnamefont{Lem{\'e}e}},
  \bibnamefont{and} \bibinfo{author}{\bibfnamefont{A.~L.}
  \bibnamefont{Goodwin}}, \bibinfo{journal}{Acta Crystallographica Section B:
  Structural Science, Crystal Engineering and Materials}
  \textbf{\bibinfo{volume}{79}}, \bibinfo{pages}{138}
  (\bibinfo{year}{2023}{\natexlab{b}}).

\bibitem[{\citenamefont{Pasciak and Welberry}(2011)}]{pasciak2011diffuse}
\bibinfo{author}{\bibfnamefont{M.}~\bibnamefont{Pasciak}} \bibnamefont{and}
  \bibinfo{author}{\bibfnamefont{T.~R.} \bibnamefont{Welberry}},
  \bibinfo{journal}{Zeitschrift für Kristallographie - Crystalline Materials}
  \textbf{\bibinfo{volume}{226}}, \bibinfo{pages}{113} (\bibinfo{year}{2011}).

\bibitem[{\citenamefont{Welberry and
  Butler}(1994)}]{welberry1994interpretation}
\bibinfo{author}{\bibfnamefont{T.}~\bibnamefont{Welberry}} \bibnamefont{and}
  \bibinfo{author}{\bibfnamefont{B.}~\bibnamefont{Butler}},
  \bibinfo{journal}{Journal of applied crystallography}
  \textbf{\bibinfo{volume}{27}}, \bibinfo{pages}{205} (\bibinfo{year}{1994}).

\end{thebibliography}
	
\end{document}